\documentclass[fleqn,usenatbib,useAMS]{mnras}

\usepackage{newtxtext,newtxmath} 

\usepackage[T1]{fontenc}
\DeclareRobustCommand{\VAN}[3]{#2}
\let\VANthebibliography\thebibliography
\def\thebibliography{\DeclareRobustCommand{\VAN}[3]{##3}\VANthebibliography}

\usepackage{amsmath}	
\usepackage{graphicx}
\usepackage{subcaption}
\usepackage{sirimacro}
\usepackage{xcolor}
\usepackage{multirow}
\usepackage{cancel}
\usepackage[normalem]{ulem}
\def\simpropto{\lower.2ex\hbox{$\; \buildrel \propto \over \sim \;$}}
\def\ltsim{\lower.5ex\hbox{$\; \buildrel < \over \sim \;$}}
\def\gtsim{\lower.5ex\hbox{$\; \buildrel > \over \sim \;$}}

\definecolor{dark blue}{rgb}{0.00, 0.00, 0.55}
\definecolor{dark green}{rgb}{0.00, 0.39, 0.00}
\definecolor{dark red}{rgb}{0.55, 0.00, 0.00}

\newcommand{\eref}[1]{equation~(\ref{#1})}
\newcommand{\esref}[1]{equations~(\ref{#1})}
\newcommand{\fref}[1]{Fig.~\ref{#1}}
\newcommand{\fsref}[2]{Fig.~\ref{#1} and~\ref{#2}}

\newcommand{\sref}[1]{Section~\ref{#1}}
\newcommand{\tref}[1]{Table~\ref{#1}}

\title{The brightest X-ray AGNs at redshift $3\lesssim z \lesssim 6$} 

\author[C. Heather et al.]{
Cameron Heather,$^{1}$\thanks{E-mail:\href{mailto:cameron.heather@warwick.ac.uk}{cameron.heather@warwick.ac.uk}}
Teeraparb Chantavat,$^2$
Siri Chongchitnan,$^{1}$
and Poemwai Chainakun$^{3,4}$
\\
$^{1}$Warwick Mathematics Institute, University of Warwick, Zeeman Building, Coventry CV4 7AL, UK\\
$^{2}$Institute for Fundamental Study, Naresuan University, Phitsanulok 65000, Thailand\\
$^{3}$School of Physics, Institute of Science, Suranaree University of Technology, Nakhon Ratchasima 30000, Thailand \\
$^{4}$Centre of Excellence in High Energy Physics and Astrophysics, Suranaree University of Technology, Nakhon Ratchasima 30000, Thailand \\
}

\date{Accepted XXX. Received YYY; in original form ZZZ}

\pubyear{2025}

\begin{document}
\label{firstpage}
\pagerange{\pageref{firstpage}--\pageref{lastpage}}
\maketitle

\begin{abstract}
 Given recent X-ray observations of high-redshift active galactic nuclei (AGNs), we consider whether the extreme luminosities of these AGNs are consistent with current semi-analytical models.  In particular, we apply extreme-value statistics (EVS) to obtain predictions of extreme X-ray luminosities of AGNs in the redshift range $3\lesssim z\lesssim 6$.  We apply this formalism using different X-ray luminosity functions and compare the predicted extreme luminosities to AGNs in the Stripe 82 X-ray catalogue. We find a general consistency between data and the EVS predictions although there is some tension with certain luminosity functions. We discuss possible extensions to this model, including extrapolating our results to even higher redshifts ($z\gtrsim10$) where AGNs have recently been observed.
\end{abstract}


\section{Introduction}
\label{sec:introduction}

Active Galactic Nuclei (AGNs) are some of the most energetic astrophysical objects in the Universe. They are created by the accretion of matter onto supermassive black holes (SMBHs) with typical masses of $M_{\rm BH} \sim 10^5$--$10^9$~${\rm M}_{\odot}$ at their cores \citep{Rees1984, Fabian2012, Yuan_Narayan2014, Netzer2015, Inayoshi_ea2020}. Among the various wavelengths used to study AGNs, X-ray observations provide a unique window into the innermost regions closest to the event horizon of the black hole.

During the accretion process, the gravitational energy of the infalling gas is converted into radiation, primarily in the form of optical and UV photons. These photons are Compton up-scattered by high-energy electrons within the corona, producing a continuum of X-ray emission whose spectrum is characterized by a power-law shape with a photon index $\Gamma$ and a cut-off energy $E_{\rm cut}$ \citep{Pozdnyakov1983, Rybicki1986, Mushotzky1993}. The AGN coronal temperature is controlled by annihilation and pair production through heating and cooling processes \citep[e.g.][]{Fabian2015}.
%


The most luminous AGNs, with 2--10 keV X-ray luminosity $L_{\rm X}$ as large as  $10^{45}$~erg~s$^{-1}$, stand out as extraordinary laboratories for understanding extreme astrophysical processes \citep[e.g.][]{Mateos2015, Stanley2017, Veronesi2023}. Their immense luminosities result from the highly efficient energy conversion in the accretion process, making them critical probes of the interplay between SMBHs and their extreme environments. 


Examining the most luminous AGNs is essential for many reasons: Their exceptional brightness enables their detection across immense cosmic distances, serving as crucial markers for investigating the high-redshift Universe \citep{Aird_ea2015}. These AGNs also act as indicators for the growth of SMBHs and galaxy evolution, spanning the epoch of reionization and beyond.  Understanding how such luminous AGNs form and evolve provides insight into the co-evolution of SMBHs and their host galaxies \citep{Kormendy_Ho2013}.  Furthermore, the extreme physical conditions within the luminous AGNs test the limits of the theoretical models. Their high accretion rates, intense radiation, and strong outflows are valuable tests of our understanding of accretion physics, radiation processes, and feedback mechanisms at extreme energies \citep{Fabian2012}.

The most luminous AGNs also play a pivotal role in shaping their cosmic environments. Their energetic feedback can regulate star formation in their host galaxies through powerful winds and jets, profoundly influencing galaxy evolution \citep{Sijacki_ea2007, Booth_Schaye2009}. At the largest scales, these AGNs impact the intergalactic medium, contributing to heating and metal enrichment \citep{Fabjan_ea2010}. By studying the most luminous AGNs, we gain a better understanding of the mechanisms by which energy is redistributed across scales, connecting SMBHs  to other larger-scale structures.

In this work, we will study the abundance and brightness of the most luminous X-ray AGNs using extreme-value statistics (EVS). This framework has previously been applied to estimate the abundances of the most massive Pop III stars (\cite{Chantavat_ea2023}), the most massive galaxy clusters \citep{chongchitnan2012primordial}, extreme primordial black holes \citep{Chongchitnan_ea2021, chongchitnanspin}, and most recently, the brightest {\it JWST} galaxies at $z\gtrsim9$ \citep{heather}. 

Our EVS modelling will allow us to test various semi-analytical models of the X-ray luminosity functions by comparing their predictions of extreme X-ray luminosities against data from the Stripe 82 X-ray catalogue, a benchmark dataset for X-ray surveys combining data from the \ii{Chandra} and \ii{XMM-Newton} space telescopes \citep{LaMassa2013, LaMassa2016, LaMassa2024}. In particular, we will use the so-called S82-XL data from \cite{Peca2024} and \cite{LaMassa2024}, and focus on AGNs at redshifts $3\lesssim z\lesssim6$. The central goal of this work is to quantify the consistency (or tension) between theory and observation. The EVS formalism does not require a complete census of all AGNs between $3<z<6$ in Stripe 82X.  Instead, we only require the X-ray luminosities of the most luminous AGNs observed in \cite{LaMassa2024} and \cite{Peca2024}. In other words, our focus is on comparing the brightest, confirmed AGNs observed in Stripe 82X to the upper bound of AGN luminosities predicted by EVS.



The rest of this paper is organized as follows. In \sref{sec:lum_funcs}, we present a survey of models of the X-ray luminosity functions and their associated observables.  In \sref{sec:agnnumber}, we present the formalism used to calculate AGN number counts. \sref{sec:extremevaluemodelling} details our EVS modelling, with the main results (the predictions of of extreme AGN luminosities across redshifts) shown in \sref{sec:results}. We discuss the implications of our findings in \sref{sec:discussionandconclusion} .

We will assume a flat $\Lambda\rm{CDM}$ cosmology with $H_0 = 70\, \rm{km\ s^{-1}\ Mpc^{-1}}$, $\Omega_{\rm m} = 0.3$ and $\Omega_{\Lambda} = 0.7$. We work in natural units with $c = 1$.

\section{The X-Ray Luminosity Function}
\label{sec:lum_funcs}

\begin{figure*}
    \centering
    \includegraphics[width=\linewidth]{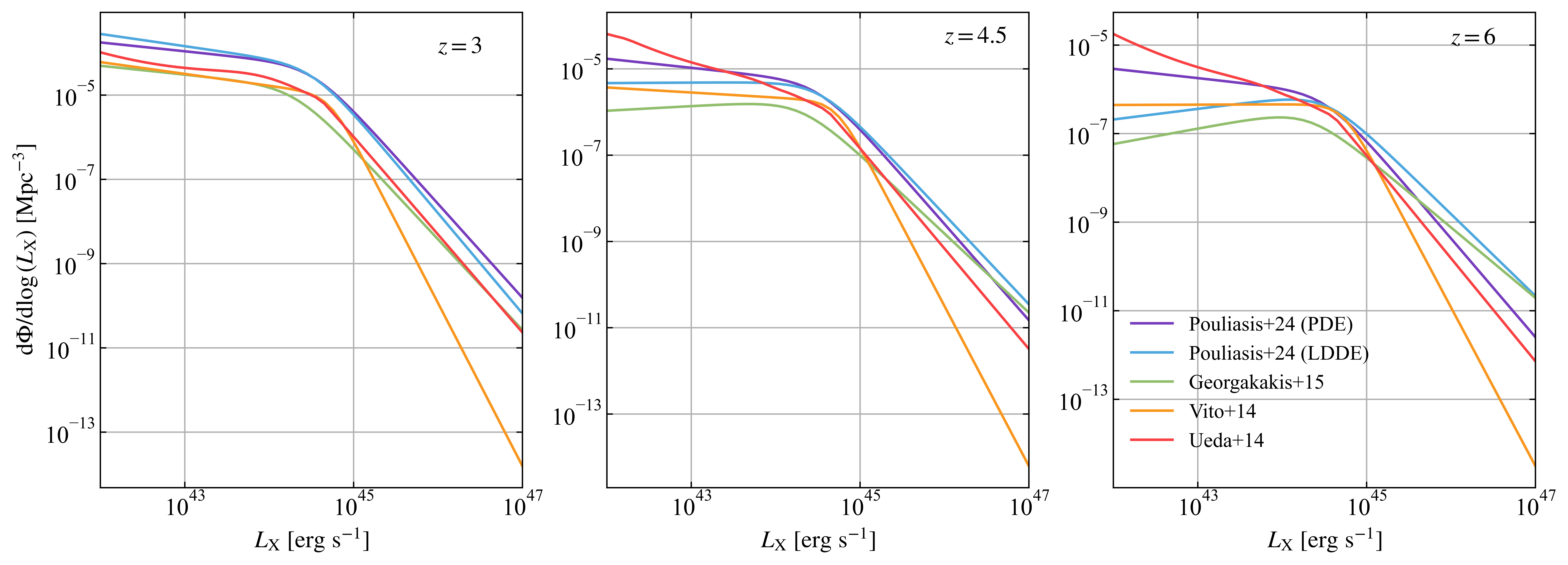}
    \caption{Comparison of the X-ray luminosity functions at $z=3$ (left), $z=4.5$ (middle) and $z=6$ (right). We include the PDE and LDDE model from \protect\cite{Pouliasis2024}, the LDDE models from \protect\cite{Georgakakis2015} and \protect\cite{Vito14}, a LDDE with a redshift cut-off from \protect\cite{Ueda14}, and the \protect\cite{Peca2023} model derived from numerical fit to observation.}
    \label{fig:lum_funcs}
\end{figure*}

The X-ray luminosity function, $\text{d}\Phi/\text{d}\log L_{\rm X}$, refers to the differential AGN number density, $\Phi$, per logarithmic interval in X-ray luminosity $L_{\rm X}$. In the literature, the luminosity function is often modelled as a double power-law function \citep{maccacaro83, maccacaro91}
\begin{equation}\label{eq:lum_z0}
    \frac{\D\Phi(L_{\rm X}, z = 0)}{\D\log L_{\rm X}} = A\left[\left(\frac{L_{\rm X}}{L_*}\right)^{\gamma_1} + \left(\frac{L_{\rm X}}{L_*}\right)^{\gamma_2}\right]^{-1},
\end{equation}
where $A$ is a normalisation factor, $L_*$ is the characteristic luminosity break, and $\gamma_1$ and $\gamma_2$ describe the slopes of the two power laws. 

To model the redshift-dependence of the luminosity function, an evolution factor, $e(z)$, is introduced, so that the X-ray luminosity function becomes
\begin{equation}
    \frac{\D\Phi(L_{\rm X}, z)}{\D \log L_{\rm X}} = \frac{\D\Phi(L_{\rm X}, z = 0)}{\D \log L_{\rm X}} e(z).
\end{equation}

In this section, we explore two forms of $e(z)$, namely the \ii{pure density evolution} (PDE) model and the \ii{luminosity-dependent density evolution} (LDDE) model. The two models differ in their dependence on X-ray luminosity. In the PDE model, we assume that the redshift evolution of the number density is the same regardless of AGN luminosity. In contrast, the LDDE model assumes that the redshift evolution depends on the luminosity of the AGNs. Although the LDDE model includes an additional parameter, it provides a better fit to the data, and its extrapolation below the survey flux limit remains consistent with the cosmic X-ray background \citep{miyaji}. We also examine a form of the LDDE model with a cut-off redshift value.   \fref{fig:lum_funcs} gives a graphical summary of the various luminosity functions  explored in this section.

\subsection{Pure density evolution (PDE) model}
\label{ssec:pde_model}

This model uses the redshift evolution factor of the form  \citep{hasinger} 
\begin{equation}
    \label{eq:ez_pde}
    e(z) = \left(\frac{1+z}{1+z_c}\right)^{p_{\rm den}}.
\end{equation}
 The parameter $p_{\rm den}$ gives the slope of the power law dependence on redshift $z$. $z_c$ is the critical redshift.
Fitting this to AGN data at redshifts $3$ -- $6$, \cite{Pouliasis2024} gave the value  $p_{\rm den} = -7.35^{+0.39}_{-0.42}$ with $z_c = 3$. Assuming the PDE model, the parameter values in the luminosity function \eref{eq:lum_z0} are 

\noindent$\displaystyle \log (A/{\text{Mpc}^{-3}}) = -4.28^{+0.09}_{-0.09}$, \quad $\log (L_*/{\text{erg  s}^{-1}}) = 44.52^{+0.07}_{-0.07}$

\noindent $\gamma_1 = 0.21^{+0.08}_{-0.10}$, \ff and \ff $\gamma_2 = 2.23^{+0.14}_{-0.13}$. 

\subsection{Luminosity-dependent density evolution (LDDE) model}
\label{ssec:ldde_model}

A generalisation of the PDE model proposed by \cite{miyaji} is the luminosity-dependent density model where
\begin{equation}
    e(z, L_{\rm X}) = \left(\frac{1+z}{1+z_c}\right)^{p_{\rm den}+\beta(\log L_{\rm X} - 44)}.
\end{equation}
$\beta$ is the slope of the luminosity-dependence redshift evolution while $p_{\rm den}$ is the slope of the power law redshift evolution similar to \eref{eq:ez_pde}.  We will consider various parametrisations of this form given in \cite{Pouliasis2024}, \cite{Georgakakis2015} and \cite{Vito14}. The associated parameter values from these references are shown in \tref{tab:LDDE}.

\begin{table}
\begin{tabular}{p{0.11\textwidth}p{0.09\textwidth}p{0.09\textwidth}p{0.09\textwidth}}
    \hline
    Parameters & \hfil Pouliasis & \hfil Georgakakis &\hfil Vito \\
    \hline
    \hfil $\log (A/{\text{Mpc}^{-3}})$ & \hfil $\displaystyle  -4.28^{+0.09}_{-0.10}$ & \hfil $\displaystyle -4.79^{+0.14}_{-0.17}$ & \hfil $\displaystyle -4.98^{+0.04}_{-0.04}$ \\[1ex]
    \hfil $\log (L_{*}/\text{erg } \text{s}^{-1})$ & \hfil $\displaystyle 44.52^{+0.07}_{-0.07}$ & \hfil $\displaystyle 44.31^{+0.13}_{-0.11}$ & \hfil $\displaystyle 44.72^{+0.09}_{-0.11}$ \\[1ex]
    \hfil $\gamma_1$ & \hfil $\displaystyle 0.29^{+0.08}_{-0.10}$ & \hfil $\displaystyle 0.21^{+0.15}_{-0.13}$ & \hfil $\displaystyle 0.28^{+0.16}_{-0.19}$ \\[1ex]
    \hfil $\gamma_2$ & \hfil $\displaystyle 2.38^{+0.15}_{-0.14}$ & \hfil $\displaystyle 2.15^{+0.24}_{-0.21}$ & \hfil $\displaystyle 3.87^{+1.08}_{-0.88}$ \\[1ex]
    \hfil $p_{\rm den}$ & \hfil $\displaystyle -8.53^{+0.58}_{-0.65}$ & \hfil $\displaystyle -7.46^{+1.03}_{-1.12}$ & \hfil $\displaystyle -6.43^{+1.12}_{-1.17}$ \\[1ex]
    \hfil $\beta$ & \hfil $\displaystyle 2.18^{+0.79}_{-0.81}$ & \hfil $\displaystyle 2.30^{+1.60}_{-1.53}$ & \hfil $\displaystyle 1.18^{+2.06}_{-2.00}$ \\[1ex]
    \hline
\end{tabular}
\caption{Parameters for the LDDE model of the X-ray luminosity function, with values from \protect\cite{Pouliasis2024}, \protect\cite{Georgakakis2015} and \protect\cite{Vito14}.}
\label{tab:LDDE}
\end{table}

\subsection{LDDE model with redshift cut-off}
\label{ssec:ueda}

This is a model with a more complicated evolution factor of the form \citep{Ueda14}. This model introduces additional redshift cut-offs, $z_{c_1}$ and $z_{c_2}$, as well as slopes of the power law; $p_1, p_2$ and $p_3$:
\begin{equation}
    e(z, L_{\rm X}) = \begin{cases}
        (1+z)^{p_1}     & [z\leq z_{c_1}(L_{\rm X})] \\[2ex]
        (1+z_{c_1})^{p_1} \left(\displaystyle\frac{1+z}{1+z_{c_1}}\right)^{p_2} & [z_{c_1}(L_{\rm X}) < z \leq z_{c_2}]    \\[2ex] 
        (1+z_{c_1})^{p_1} \left(\displaystyle\frac{1+z}{1+z_{c_1}}\right)^{p_2}\left(\displaystyle\frac{1+z}  {1+z_{c_2}}\right)^{p_3}&[z>z_{c_2}],
    \end{cases}
\end{equation}
where $p_2 = -1.5$, $p_3 = -6.2$ and $p_1$ is given by the following function of luminosity
\begin{equation}
    p_1(L_{\rm X}) = p_1^* + \beta_1 \log\bkt{L_{\rm X}/44},
\end{equation}
with $p_1^* =4.78^{+ 0.16}_{-0.16}$, $\beta_1 = 0.84^{+0.18}_{-0.18}$. Assuming this model, the parameter values in the luminosity function \eref{eq:lum_z0} are  $ \log (A/{\text{Mpc}^{-3}}) = -5.536^{+0.01}_{-0.01}$,  $\log (L_*/\text{erg } \text{s}^{-1}) = 43.97^{+0.06}_{-0.06}$, $\gamma_1 = 0.09^{+0.04}_{-0.04}$, $\gamma_2 = 2.71^{+0.09}_{-0.09}$.

The redshift breaks $z_{c_1}$ and $z_{c_2}$ are themselves functions of luminosity, given by a power-law with cut-offs:
\begin{equation}
    z_{c_1}(L_{\rm X})= 
    \begin{cases}
        z^*_{c_1}(L_{\rm X}/L_{a_1})^{\alpha_1}& [L_{\rm X} \leq L_{a_1}]   \\[2ex]
        z^*_{c_1}   & [L_{\rm X} > L_{a_1}],
    \end{cases}
\end{equation}
and 
\begin{equation}
    z_{c_2}(L_{\rm X})= \begin{cases}
        z^*_{c_2}(L_{\rm X}/L_{a_2})^{\alpha_2} & [L_{\rm X} \leq L_{a_2}]   \\[2ex]
        z^*_{c_2}   & [L_{\rm X} > L_{a_2}].
    \end{cases}
\end{equation}
The parameters values are given in \cite{Ueda14}. We list them here for completeness: \\
$z^*_{c_1} = 1.86_{+0.07}^{-0.07}, \log (L_{a_1}/\text{erg } \text{s}^{-1}) = 44.61^{+0.07}_{-0.07}, \alpha_1 = 0.29^{+0.02}_{-0.02}$\\
$z^*_{c_2}=3$, $\log (L_{a_2}/\text{erg } \text{s}^{-1}) = 44$, $\alpha_2=-0.1$.

\subsection{Numerical fit model}
\label{ssec:numerical}

We have also included the luminosity function from \cite{Peca2023}, which is a numerical fit to the observed $L_{\rm X}$ values from the \ii{Stripe 82X} catalogue up to $z = 4$. We numerically extrapolated the results to $z = 6$ using a linear grid interpolator on the $L_{\rm X}$-$z$ plane; the data is fitted as a surface on this rectilinear grid and values outside this
domain can be extrapolated, just as the authors have done in their work.

\bigskip

\fref{fig:lum_funcs} shows the various luminosity functions at three redshift values, $z=3, 4.5$ and $6$. We choose to work with the central values for the parameters given, since the errors provided are mostly small and would minimally affect the EVS results in \sref{sec:extremevaluemodelling}.

In \fref{fig:lum_funcs}, we see that the luminosity functions are generally of the same double-power law shape with similar magnitudes. However, at the high luminosity end, the Vito model decreases most steeply, predicting the least number of the brightest AGNs. We also note that as redshift increases, the low luminosity end shows increasing disparity amongst the models, with the Ueda model dominating above other models at $z\approx6$.


\section{AGN number count}
\label{sec:agnnumber}

For an AGN with X-ray luminosity $L_{\rm X}$ at redshift $z$, we also consider its hydrogen column density $N_{\rm H}$, which quantifies how obscured the AGN is. Higher $N_{\rm H}$ values indicate significant obscuration, for example, due to absorption by a dusty torus or host-galaxy gases \citep{Antonucci_1993, Netzer2015}.  Compton-thick AGNs  with $N_{\rm H} > 10^{24}\, {\rm cm}^{-2}$ can completely obscure soft X-ray emissions.  Understanding $N_{\rm H}$ is essential for probing the physical properties of AGN environments and their impact on observed spectra \citep{Elitzur_Shlosman2006}.
 

The number count of AGNs in a given comoving volume $V$ can be written as  \citep{Vijarnwannaluk_ea2022}
\begin{eqnarray} \label{eq:NumCount}
    \nonumber N &=& \iiint f_{\rm abs}(\log L_{\rm X}, z, N_{\rm H}) \frac{\D\Phi(\log L_{\rm X},z)}{\D\log L_{\rm X}} \\
    && \times f_{\rm sky}\frac{\D V}{\D z}\D\log L_{\rm X} \ \D z\  \D \log N_{\rm H},
\end{eqnarray}
where
\begin{eqnarray}
    \frac{\D V }{\D z} = \frac{4\pi}{H(z)} \left(\int_0^z\frac{\D z'}{H(z')}\right)^2, \\
    H(z) = H_0\left[\Omega_{\rm m}(1+z)^3 + \Omega_{\Lambda}\right]^{\frac{1}{2}}.
\end{eqnarray}

The absorption function $f_{\rm abs}$ depends on X-ray luminosity, redshift and $N_{\rm H}$. While \cite{Ueda14} gave an updated XLF and absorption function built into a population synthesis model, we opted to use the functional form of $f_{\rm abs}$ from \cite{Ueda_ea2003} because it provides a simple analytic prescription that is widely used in XLF modelling, enabling a more direct comparison with prior works. The differences between both models are minimal at the high X-ray luminosities that dominate our samples. The Compton-thick AGNs ($N_{\rm H} > 10^{24}~{\rm cm}^{-2}$) which are not explicitly included in our $f_{\rm abs}$ integration, but are more carefully treated in \cite{Ueda14}, are  not included in our sample. In particular, there are no Compton-thick AGN reported in the catalogue in the redshift range $3\leq z\leq6$. Indeed, the observed Stripe 82X catalogue showed a median $N_{\rm H} = 10^{21.6}~{\rm cm}^{-2}$, with only ~37\% of sources having $N_{\rm H} > 10^{22}~{\rm cm}^{-2}$ \citep{Peca2024}. In other words, most of the brightest AGNs in our sample are essentially unobscured, so omitting Compton-thick AGNs is expected to have only a negligible effect on our results.

The number count in \eref{eq:NumCount} requires \ii{intrinsic} X-ray luminosity, whereas the AGN from our sample are given in observed X-ray luminosity. We find the difference between the observed and intrinsic luminosity to be negligible, as verified using PIMMS\footnote{\url{https://heasarc.gsfc.nasa.gov/docs/software/tools/pimms.html}} under the standard spectral assumptions for unobscured AGNs (e.g. $\Gamma = 1.8$). We found that even at the higher column densities in our sample ($N_{\rm H} \approx  10^{23.5} \rm \ cm^{-2}$), the percentage difference between the intrinsic and observed luminosities is still relatively small ($\sim  10\%$ for $z = 3$, decreasing to $\sim 2.5\%$ for $z = 6$).
 The percentage difference could be higher for Compton-thick AGNs but these are not present in our data.

The S82-XL survey reaches very faint X-ray fluxes (hard-band 2--10 keV limit $\sim 2.9 \times 10^{-15}$~erg~s$^{-1}$~cm$^{-2}$).
We found that  34 of our 35 sources are above the flux limit, with almost all having a flux greater than $10^{14} \rm erg\ s^{-1}\ cm^{-2}$. There is only one source we are using that is below this flux limit, which corresponds to the AGN at $z = 5.855$, previously studied elsewhere \cite{Paris09}. We chose to keep this point as it has the highest spectroscopically confirmed redshift in the catalogue, which is important in our EVS discussion.

Although the sky coverage can vary with flux sensitivity (introducing a bias against Compton-thick AGNs, which appear fainter), this effect is negligible for our sample of bright AGNs, the majority of which are also Compton-thin and detectable across the entire survey area. We therefore adopt a constant $f_{\rm sky}=54.8$~deg$^2$ in our analysis, as in \cite{Peca2024}. Other factors, such as the incompleteness of spectroscopic redshifts, may also directly affect the expected maximum luminosity from the XLF models. We discuss an uncertainty factor, $f_{\rm unc}$, which modulate the number counts, in Appendix~\ref{appendix_uncertainties}. In any case, we find that this has a subdominant effect on our extreme-value distributions and does not change our main conclusions.


%
\section{Extreme-Value Modelling of the Brightest AGNs}
\label{sec:extremevaluemodelling}

\subsection{Extreme-value statistics}
\label{ssec:evs}

\begin{figure*}
    \centering
    \includegraphics[width=\linewidth]{paper_plots/pdf_plots.png}
    \caption{The probability density functions for the most luminous AGNs calculated in \sref{sec:lum_funcs} for three redshift bins: $ z = [3,3.5]$ (left), $z = [4.5, 5]$ (middle) and $z = [5, 5.5]$ (right). These plots show the EVS pdf taken to be Gumbel, where we assume $\gamma = 0$. The values for $\alpha$ and $\beta$ which describe the curves are given in \tref{tab:full}.}
    \label{fig:pdf_plots}
\end{figure*}

Here we give an overview of how the EVS framework can be used to determine the luminosity of the brightest X-ray AGNs expected in a given volume. 

We utilise the generalised extreme-value (GEV) method, also known as the block maxima method \citep{gumbel1958statistics, deHaan2006}. In this method, we divide the AGN population in a given redshift bin into $S$ distinct blocks, from each of which we identify the brightest AGN. Analogous to the Central Limit Theorem, in the large-$S$ limit, the distribution of the brightest AGNs will approach the generalised extreme-value distribution (\eref{eq:GEV}). 

The calculation pipeline is as follows. First, we calculate the number count of $N(>L_{\rm X})$ of AGNs with X-ray luminosity greater than $L_{\rm X}$ in a given redshift bin using \eref{eq:NumCount}. Next, we consider the probability $P_0$ that no AGNs in a given volume exceeds the maximum luminosity $L_{\rm X}$. The probability distribution $P_0(L_{\rm X})$ can be modelled as a Poisson distribution with the following cumulative distribution function (cdf):
\begin{eqnarray}\label{eq:Poisson}
    P_0(L_{\rm X}) = \exp(-N(>L_{\rm X})),
\end{eqnarray}
Differentiating this with respect to $L_{\rm X}$, we obtain the probability density function (pdf):
\begin{eqnarray}\label{eq:Pois_pdf}
   \frac{\D P_0}{\D L_{\rm X}} = -\frac{\D N(>L_{\rm X})}{\D L_{\rm X}}P_0(L_{\rm X}),
\end{eqnarray}

 The  Fisher-Tippett-Gnedenko theorem implies  that in the large-$N$ limit, the cdf (\eref{eq:Poisson}) approaches the GEV distribution given by:
\begin{eqnarray}
    G(L_{\rm X}) = 
    \begin{cases}
        \exp\left[ -(1+\gamma y)^{-1/\gamma}\right] & (\gamma \neq 0), \\
        \exp\left[ -e^{-y}\right] & (\gamma = 0),
    \end{cases}\label{eq:GEV}
\end{eqnarray}
where $ y = (L_{\rm X} - \alpha)/ \beta$, with $\alpha$ describing the location of the peak, and $\beta$ describing the scale of the pdf. The sign of the parameter $\gamma$ determines the GEV type, with $\gamma = 0$, $\gamma>0$ and $\gamma<0,$ corresponding to the Gumbel, Fr\'echet and Weibull distributions respectively. 

We can express the GEV parameters $\alpha$, $\beta$ and $\gamma$ in terms of astrophysical parameters by Taylor-expanding the Poisson distribution and GEV distribution around the peak of the pdf at $L_{\rm X}$ to cubic order. By equating coefficients, we find

\begin{eqnarray}\label{eq:params}
\begin{aligned}
    \gamma &= N(>L_{\rm peak}) - 1, \\\\
    \alpha & = L_{\rm peak} - \frac{\beta}{\gamma}\left({(1+\gamma)}^{-\gamma}-1\right), 
\end{aligned} 
\begin{aligned}
    \beta &= \frac{{(1+\gamma)}^{1+\gamma}}{\displaystyle\frac{\D N}{\D L_{\rm X}}\Big|_{L_{\rm peak}}}.
\end{aligned}
\end{eqnarray}

tIn \fref{fig:pdf_plots}, we plot the extreme-value pdf for the brightest AGNs for three redshift bins $ z = [3, 3.5], [4.5, 5]$ and $[5, 5.5]$ where the distributions shown are calculated from \eref{eq:GEV}. The values for $\alpha$ and $\beta$, are calculated from \esref{eq:params} and shown in \tref{tab:full} in the Appendix B. We note from \tref{tab:full} that $\alpha$ is indeed close to the peak luminosity in each redshift bin.

\begin{figure}
    \centering
    \includegraphics[width=\linewidth]{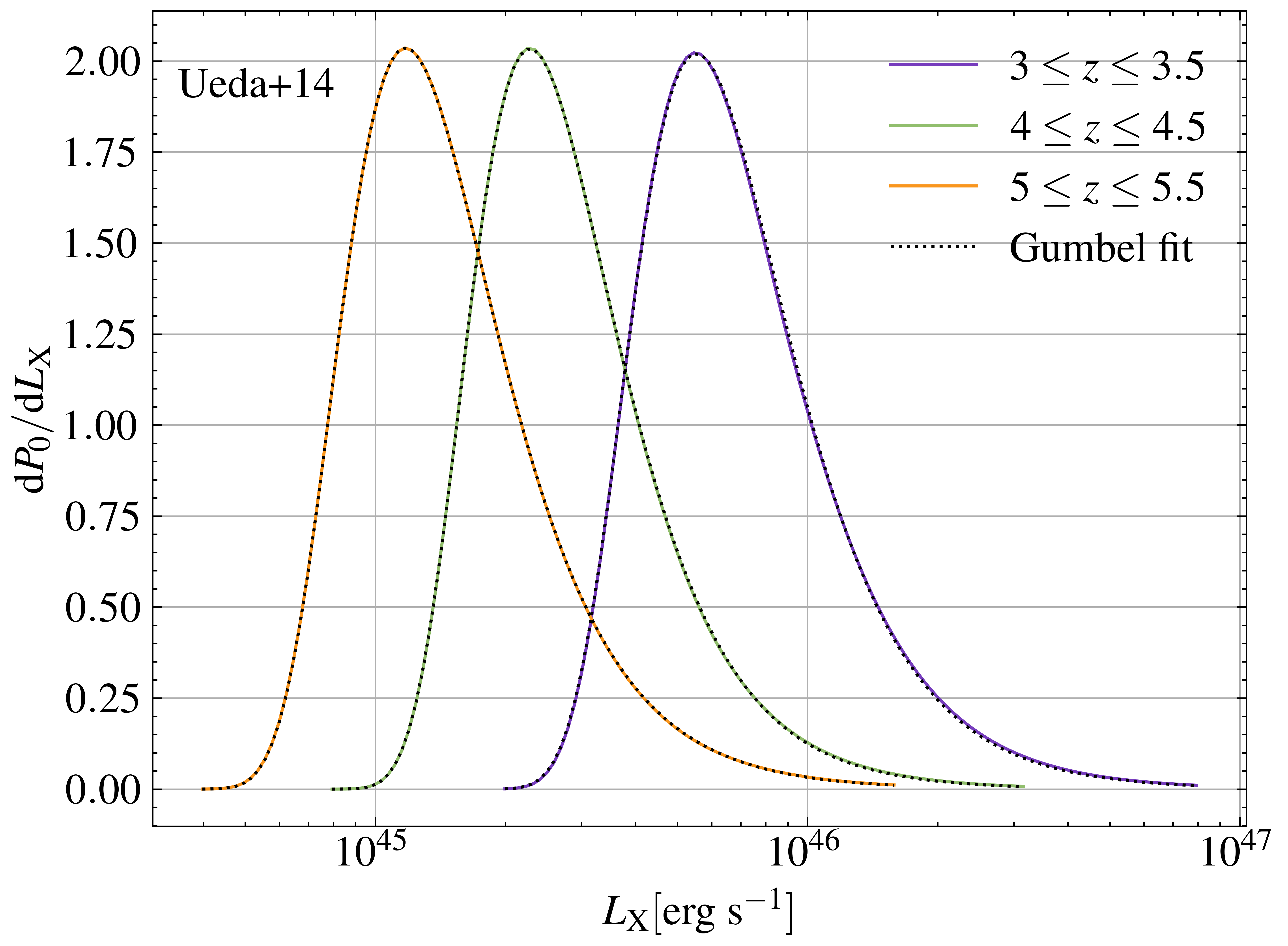}
    \caption{A plot to show the comparison of the Gumbel fit to the Poisson pdf with three redshift bins for the cut-off redshift LDDE luminosity function in \sref{ssec:ueda}. The solid coloured lines show the pdf derived from the Poisson distribution (\eref{eq:Pois_pdf}) and the dashed lines are the Gumbel distribution (\eref{eq:GEV}) with $\gamma =  0$ and the values for $\alpha$ and $\beta$ given in \tref{tab:full}.  We can see that choosing the Gumbel distribution, where we take $\gamma = 0$, is a suitable fit for our data. }\label{fig:ueda_pdf}
\end{figure}

We found that for each of the luminosity functions, the value of $\gamma$ is small ($\gamma \lesssim |0.05|$). Hence, the GEV distribution is well approximated by the Gumbel distribution $(\gamma=0)$. In \fref{fig:ueda_pdf}, we give a visual comparison of the extreme-value pdf calculated from \eref{eq:Pois_pdf} to its Gumbel approximation, using the \cite{Ueda14} model as an example. 

From these figures, we see that as redshift increases, the peak extreme values of $L_{\rm X}$ decrease. We also see that different luminosity functions decrease at different rates. For example, the PDE model from \cite{Pouliasis2024} decreases in luminosity at a higher rate than its  LDDE variation.


Our next goal is to see whether any of these extreme-value pdfs are consistent with observation of the brightest X-ray AGNs.

\subsection{\ii{Stripe 82} X-ray AGN data}
\label{ssec:s82}

The Stripe 82 X-ray survey is a comprehensive astronomical survey focused on detecting rare, high-luminosity AGNs and other X-ray sources within the area covered by the \ii{Stripe 82} field of the Sloan Digital Sky Survey (SDSS). The survey incorporates data from the {\it XMM-Newton} and {\it Chandra} telescopes to create a comprehensive catalogue of X-ray sources \citep[e.g.][]{LaMassa2013, LaMassa2016}. The continuous updates to the catalogue have been invaluable to the study of the high-redshift Universe.

We take data from the \ii{S82X} catalogue as described in  \cite{Peca2024}. This catalogues provides samples of AGN at redshift values  $3 \lesssim z \lesssim 6$, with a sky survey area of 54.8~deg$^2$. We are using the 2--10 keV observed luminosity from the catalogue as a proxy for $L_{\rm X}$.

For the EVS analysis, we use a subset of  35 most luminous AGNs from the S82X data from \cite{Peca2024} that were in the redshift range $3 \lesssim z \lesssim 6$, and were above a certain threshold luminosity. In the catalogue there is a large number of sources with redshift $3 \lesssim z \lesssim 3.5$ with luminosity $ 44 \lesssim \log (L_{\rm X}) \lesssim 45 $. We chose to restrict the number of sources to only the brightest ones in this redshift range.




 \begin{figure*}
    \centering
    \includegraphics[width=\linewidth]{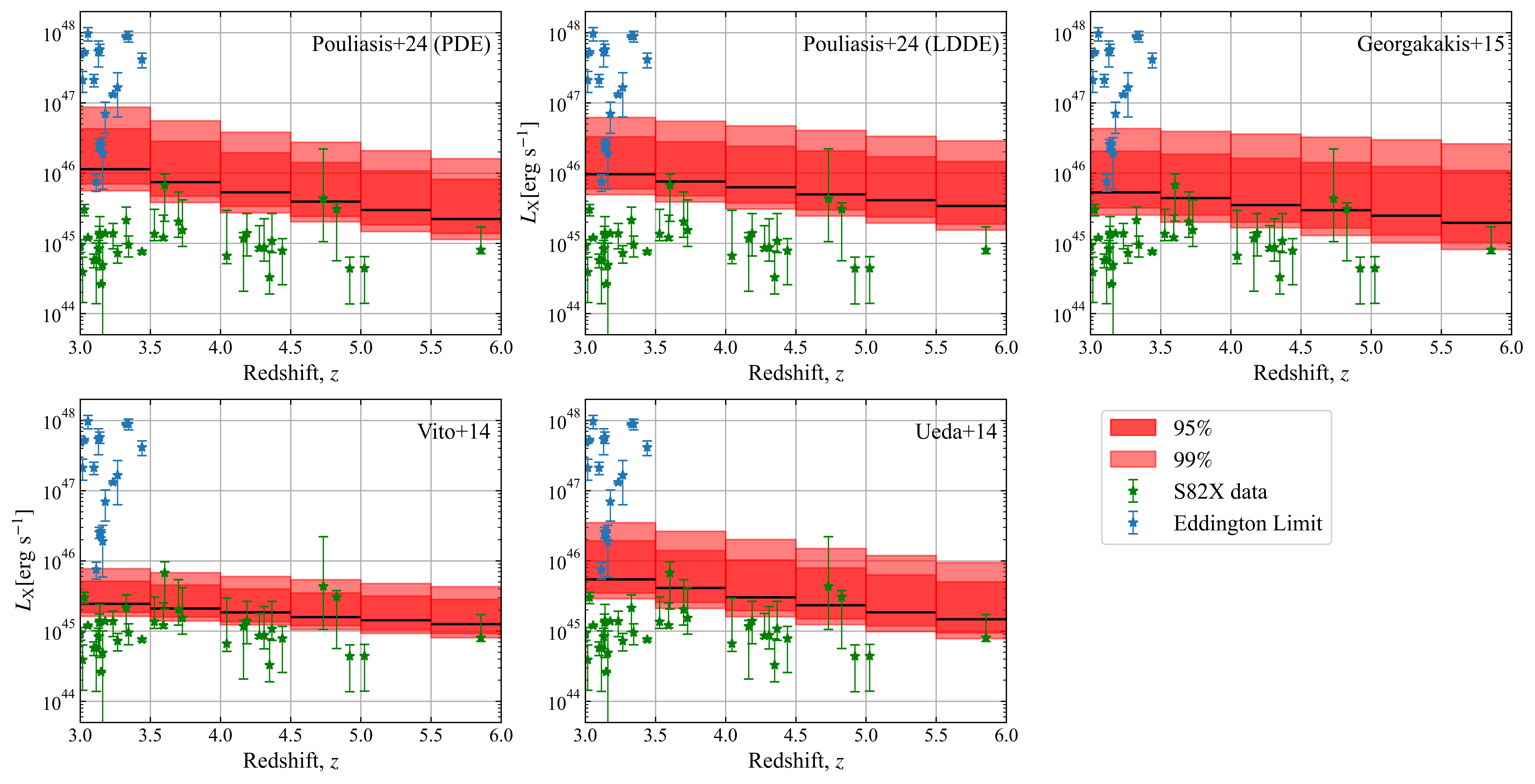}
    \caption{The peak luminosities of $L_{\rm X}$ (solid black lines) of the extreme-value (Gumbel) pdf for the $L_{\rm X}$ functions from \sref{sec:lum_funcs}, plotted in redshift bins of size $0.5$ from $z =3$ to $z = 6$. Each block shows a profile of the EVS probability density function shown in \fref{fig:pdf_plots} where the darker/ lighter shaded regions correspond to the $95^{th}$ and $99^{th}$ percentiles. The data points (green) present the most luminous AGN in the \ii{S82-XL} survey from \protect\cite{Peca2024}.}
    \label{fig:EVS_blocks}
\end{figure*}
\section{Results}
\label{sec:results}

The main result of this work is shown in \fref{fig:EVS_blocks}. The figure shows the profile of the extreme-value pdfs (showing the peak values in black, plus the 95th and 99th percentile bands) across redshift bins in the range $3\leq z\leq 6$, assuming five different luminosity functions  discussed in \sref{sec:lum_funcs}. The step-like feature comes from the fact that we calculate a single EVS distribution in each redshift bin of width 0.5. We include data points from the \ii{S82X} catalogue described in \sref{ssec:s82}. The selection criteria described in the previous section explain why our sample size in \fref{fig:EVS_blocks} appears to be smaller than those reported in  \cite{Peca2024}.


Overall we see general consistency between the brightest \ii{S82X} data and the EVS predictions, and note the following.
\bit
\item We see three AGNs that seem to be in slight tension with the EVS predictions of extreme luminosities. These are \ii{2CXO J001217.1-005437} and \ii{2CXO J021043.1-001817} from \ii{Chandra}, and \ii{4XMM J000748.9+004119} from \ii{XMM-Newton}, with redshifts $z \approx 3.60, 4.73, 4.83$. respectively. For certain luminosity functions (\cite{Georgakakis2015}, \cite{Vito14}, \cite{Ueda14}, \cite{Peca2023}), these AGNs lie above the expected peak values.

\item The \cite{Vito14} model has the narrowest EVS bands, centering around lower values of $L_{\rm X}$ than other models (\ie\ the values of the EVS parameters $\alpha$ and $\beta$ in Table \ref{tab:full} are the smallest amongst the models). The brightest AGNs straddle the outer edges of this band, showing some tension. 
The hard X-ray AGN luminosity function in \cite{Vito14} may be biased due to its calibration from soft-band X-ray AGN samples and assuming a simple power-law spectrum with photon index $\Gamma=1.8$. This could explain the sharp drop in $L_{\rm X}$ at high luminosities in \fref{fig:lum_funcs} in contrast with other luminosity functions.


\item The  \cite{Georgakakis2015} model has the broadest EVS band, and share a similar shape with the \cite{Pouliasis2024} LDDE model. Their extreme luminosity values show a more gradual decline with redshift. 

\item  The  \cite{Pouliasis2024} (PDE), the \cite{Ueda14} and the \cite{Peca2023} models have similar profiles, with the PDE model predicting slightly higher values for extreme luminosity over the redshift range.


\eit


\section{Discussion and conclusion}
\label{sec:discussionandconclusion}

Using the EVS formalism, we obtained the prediction for the extreme X-ray luminosities of AGNs in the redshift range $3 \lesssim z \lesssim 6$. We derived the probability distribution of extreme luminosities, which we found to be well approximated by the Gumbel distribution. Our main results are shown in \fsref{fig:EVS_blocks}{fig:EVS_lines}, where we compare observational data with our EVS predictions assuming 5 different semi-analytic models of the luminosity functions. We find good agreement between the models and data, with the Vito model showing most tension with data.

While there might be some biases from survey sensitivity, completeness, and absorption distribution, these factors are largely not critical to our specific analysis. For the brightest AGNs whose fluxes are well above the \ii{S82X} flux limits, biases related to absorption and survey sensitivity have minimal impact on the interpretation of our EVS results. Future studies could test this assumption, for example, by using absorption-corrected luminosities where available. Nevertheless, our study does not aim to model the entire underlying AGN population, nor do we attempt to constrain the full AGN XLF. Instead, our focus is on comparing the brightest, confirmed AGN observed in \ii{S82X} to the upper envelope of AGN luminosities predicted by extreme value statistics. One of our key results is that the brightest X-ray luminosities in \ii{S82X} mostly fall below the theoretical maximum predicted by EVS assuming various LF models. This conclusion holds true regardless of the number of undetected or excluded fainter AGNs.

Further discussion points are as follows.

\begin{itemize}
\item \ii{At lower redshifts}: \cite{Elias-Chavez2024} analysed the X-ray properties of 23 brightest AGNs from the {\it XMM-Newton} Ultra Narrow Deep Field survey, with redshifts up to 2.66. Their 2--10 keV luminosity was found to be $L_{\rm X} \sim 10^{42}$--$10^{46}$~erg~s$^{-1}$. Despite ranking among the most luminous AGNs observed in X-ray surveys to date, their luminosities are still consistent with the EVS predictions for all luminosity functions investigated here when extrapolated at the lower end.

\smallskip 

\item \ii{At higher redshifts}: We can extend the EVS modelling to estimate the extreme X-ray luminosities of AGNs at, say, $z \sim 10$ where an AGN observation has been made \citep{Goulding23}. 

    \begin{figure}
    \centering
    \includegraphics[width=\linewidth]{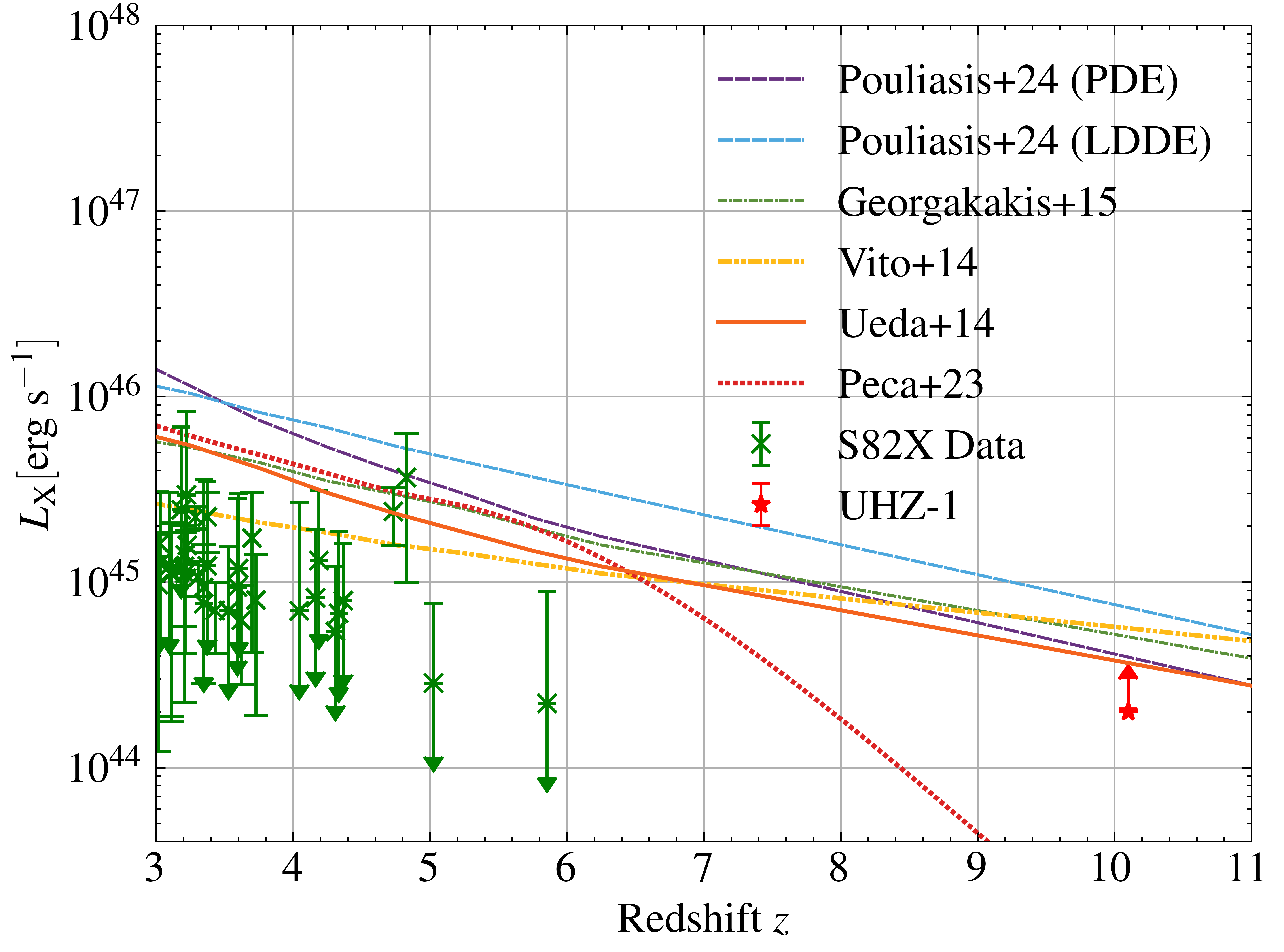}
    \caption{A comparison of EVS predictions for various luminosity functions from \sref{sec:lum_funcs}, extended to redshift $z = 11$. We include data points from \ii{S82-XL} for $z> 3$, as well as UHZ-1  \protect\citep{Goulding23} at redshift $z = 10.1$. Each line shows the peak of the extreme-value pdf for each luminosity function (percentile bands not shown).}
    \label{fig:EVS_lines}
    \end{figure}

    In \fref{fig:EVS_lines}, we plot the peak values of the extreme-value pdf, smoothed over redshifts, extended to $z=11$. We have shown the redshift range $3 < z < 11$, to avoid overcrowding of the data points in the low redshift range $3 < z <3.5$. The lines correspond to the smoothed version of each model shown in \fref{fig:EVS_blocks} up to $z=6$. We see that at $z\sim 10$, the peak of the EVS distribution is around $10^{44}-10^{45}$ erg s$^{-1}$ for most models considered in this paper with the exception of the \cite{Peca2023} model, which is most likely due to  extrapolating too far out of the redshift range used to the construction of the numerical fit.

    We also include an additional data point, UHZ-1, from \cite{Goulding23}, which has been spectroscopically confirmed at $z = 10.1$ with $L_{\rm X} \gtrsim 2 \times 10^{44}\ {\rm erg\ s^{-1}}$. Its luminosity is in agreement with the EVS prediction from various luminosity functions. 

    \smallskip
    
\item \ii{Future direction}: The luminosity functions we used have not been calibrated to these higher redshifts, where a more comprehensive semi-analytic model is needed along with more observational data. \cite{Habouzit2022} compared AGN populations across various large-scale hydrodynamical simulations and found significant discrepancies in AGN luminosity functions, especially at high redshifts. These inconsistencies reveal key challenges in our understanding of AGN feedback, obscuration patterns, duty cycle and accretion processes in the very early Universe. Future X-ray telescopes such as {\it Athena}\footnote{\url{https://www.the-athena-x-ray-observatory.eu}}, {\it AXIS}\footnote{\url{https://axis.umd.edu}}, and {\it LynX}\footnote{\url{https://www.lynxobservatory.com}} will refine these models through their enhanced sensitivity, enabling detection of faint AGNs at unprecedented redshifts and providing new insights into AGN obscuration and evolution \citep[e.g.][]{Schirra2021}.

\end{itemize}

\section*{Acknowledgements}
We would like to thank to Bovornpratch Vijarnwannaluk and Alessandro Peca for their helpful suggestions.
C.H. is supported by the Warwick Mathematics Institute Centre for Doctoral Training, and gratefully acknowledges funding from the University of Warwick and the UK Engineering and Physical Sciences Research Council (grant number: EP/W524645/1). P.C. thanks funding support from the National Science Research and Innovation Fund (NSRF) via the Program Management Unit for Human Resources \& Institutional Development, Research and Innovation [grant number B46G670083]).

\section*{Data Availability}

The data used in this article will be shared on reasonable request to the corresponding author.

\bibliographystyle{mnras} 
\bibliography{agn_main}

\appendix
\section{Uncertainties in Number Count}
\label{appendix_uncertainties}

In this section we discuss the uncertainties that could arise in the number count in equation \eqref{eq:NumCount}. We incorporate the possible uncertainties into one constant $f_{\rm unc}$ defined through the following equation:
\begin{eqnarray} \label{eq:A1}
     \nonumber N &=& f_{\rm unc}\iiint f_{\rm abs}(\log L_{\rm X}, z, N_{\rm H}) \frac{\D\Phi(\log L_{\rm X},z)}{\D\log L_{\rm X}} \\
    && \times f_{\rm sky}\frac{\D V}{\D z}\D\log L_{\rm X} \ \D z\  \D \log N_{\rm H}.
\end{eqnarray}

\begin{figure}
    \centering
    \includegraphics[width=\linewidth]{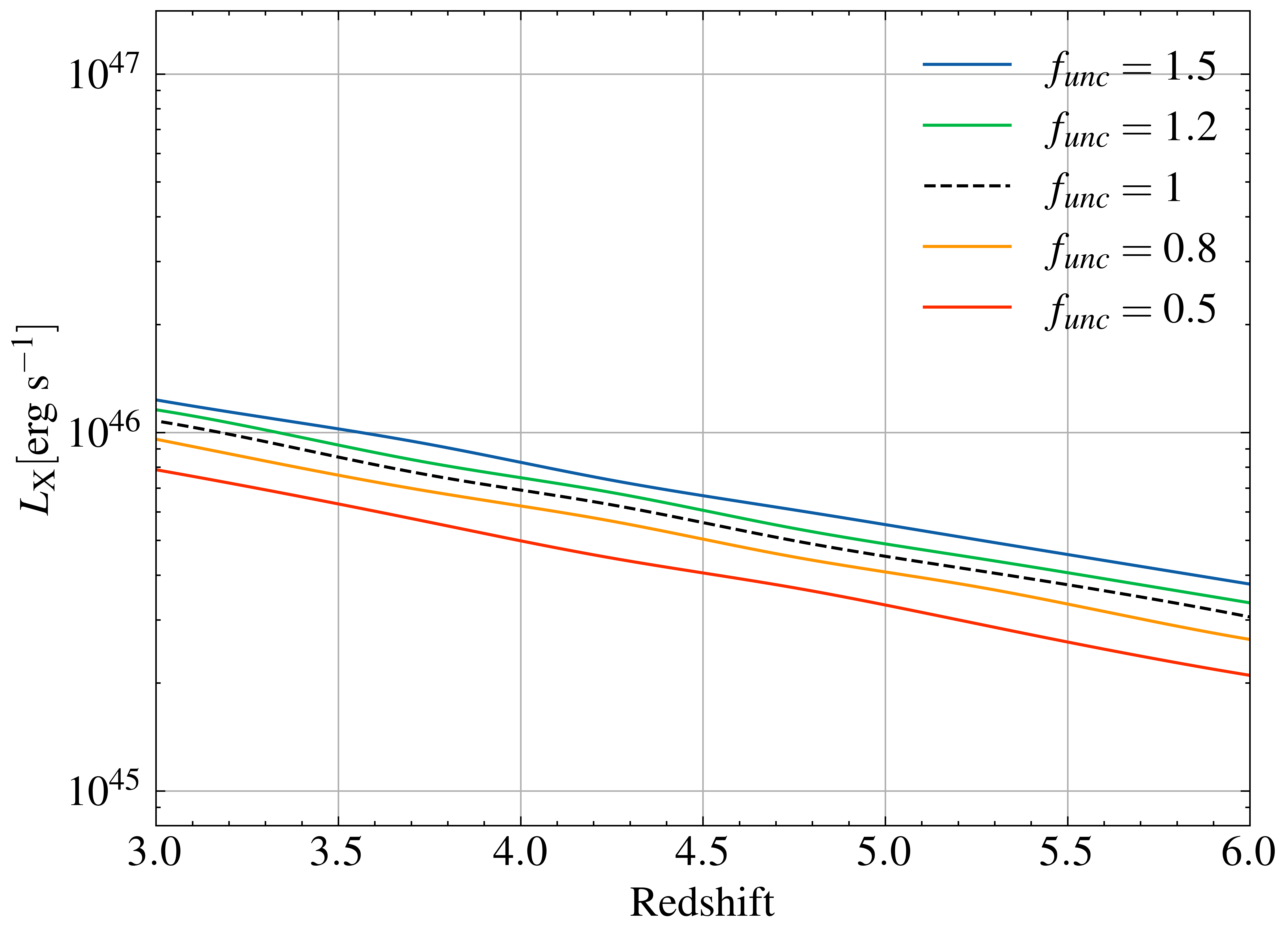}
    \caption{A comparison of EVS predictions for the \protect\cite{Pouliasis2024} PDE luminosity model when applying uncertainty factors of $f_{\rm unc} = 0.5, 0.8, 1, 1.2, 1.5$, where the black dashed line shows the peak luminosity with no uncertainty, or $f_{\rm unc} = 1$, applied to the number count defined in \eref{eq:A1}. This curve corresponds to the smoothed EVS profile shown in \fref{fig:EVS_blocks}.}
    \label{fig:uncertainties}
    \end{figure}

In \fref{fig:uncertainties} we plot the EVS predictions for $f_{\rm unc} = 0.5, 0.8, 1, 1.2, 1.5$, where $f_{\rm unc} = 1$ corresponds to no uncertainties in the number count. From this figure, we see that even for the large uncertainties, the peak luminosity changes relatively little ($\sim\pm20\%$).

We expect uncertainties from redshift incompleteness (with $f_{\rm unc} < 1$) to have negligible impact on our analysis. For instance, the $S82X$ catalogue of \cite{Peca2023} has almost $100\%$ redshift completeness, and for our subset of data points  in Fig. \ref{fig:EVS_blocks}, we only included those with spectroscopically confirmed redshifts. 


We used $f_{\rm unc} > 1$ to illustrate the opposite effects from various factors such as Malmquist bias (describing the loss of faint sources due to the flux limit in a survey) or uncertainty in the value of $f\sub{sky}$.
Both \cite{Vito14} and \cite{Georgakakis2015} include an uncertainty factor as a function of $(L_{\rm X},\ z,\ N_{\rm H})$ to account for the redshift incompleteness in the AGN samples they use.

While these uncertainties are important to acknowledge, we can see here that they do not meaningfully change our EVS calculations of peak luminosities.

\section{GEV parameters}
\label{appendix}

\tref{tab:full} shows the peak luminosities and GEV parameters ($\alpha$ and $\beta$) for the various luminosity functions used in each redshift bin, to four significant figures. The parameters have been calculated using \eref{eq:params} assuming $\gamma = 0$, in other words assuming the Gumbel distribution.

\begin{table*}
    \centering
    \begin{tabular}{||ccccc||}
        \hline
       X-ray luminosity function & Redshift & Peak $L_{\rm X}$ & $\alpha$ & $\beta$\\
        \hline
        \cite{Pouliasis2024}  & $3 \leq z \leq 3.5$ & 46.06 & 46.05 & 0.1950 \\
        (PDE) & $3.5 \leq z \leq 4$ & 45.87 & 45.88 & 0.1929 \\
        & $4 \leq z \leq 4.5$ & 45.73 & 45.73 & 0.1902 \\
        & $4.5 \leq z \leq 5$ & 45.59 & 45.59 & 0.1894 \\
        & $5 \leq z \leq 5.5$ & 45.48 & 45.47 & 0.1899 \\
        & $5.5 \leq z \leq 6$ & 45.35 & 45.35 & 0.1902 \\
        \hline
        \cite{Pouliasis2024}  & $3 \leq z \leq 3.5$ & 45.98 & 45.98 & 0.1870 \\
        (LDDE) & $3.5 \leq z \leq 4$ & 45.88 & 45.89 & 0.1934 \\
        & $4 \leq z \leq 4.5$ & 45.80 & 45.80 & 0.2002 \\
        & $4.5 \leq z \leq 5$ & 45.70 & 45.70 & 0.2075 \\
        & $5 \leq z \leq 5.5$ & 45.62 & 45.61 & 0.2149 \\
        & $5.5 \leq z \leq 6$ & 45.53 & 45.52 & 0.2231 \\
        \hline
        \cite{Georgakakis2015} & $3 \leq z \leq 3.5$ & 45.78 & 45.79 & 0.1924 \\
        & $3.5 \leq z \leq 4$ & 45.73 & 45.72 & 0.2024 \\
        & $4 \leq z \leq 4.5$ & 45.65 & 45.64 & 0.2128 \\
        & $4.5 \leq z \leq 5$ & 45.55 & 45.55 & 0.2235 \\
        & $5 \leq z \leq 5.5$ & 45.47 & 45.47 & 0.2340 \\
        & $5.5 \leq z \leq 6$ & 45.39 & 45.38 & 0.2454 \\
        \hline
        \cite{Vito14} & $3 \leq z \leq 3.5$ & 45.34 & 45.38 & 0.1111 \\
        & $3.5 \leq z \leq 4$ & 45.32 & 45.32 & 0.1128 \\
        & $4 \leq z \leq 4.5$ & 45.27 & 45.26 & 0.1149 \\
        & $4.5 \leq z \leq 5$ & 45.20 & 45.20 & 0.1166 \\
        & $5 \leq z \leq 5.5$ & 45.16 & 45.15 & 0.1187 \\
        & $5.5 \leq z \leq 6$ & 45.10 & 45.10 & 0.1206 \\
        \hline
        \cite{Ueda14} & $3 \leq z \leq 3.5$ & 45.74 & 45.74 & 0.1824 \\
        & $3.5 \leq z \leq 4$ & 45.62 & 45.61 & 0.1815 \\
        & $4 \leq z \leq 4.5$ & 45.48 & 45.49 & 0.1811 \\
        & $4.5 \leq z \leq 5$ & 45.37 & 45.37 & 0.1811  \\
        & $5 \leq z \leq 5.5$ & 45.27 & 45.27 & 0.1811 \\
        & $5.5 \leq z \leq 6$ & 45.17 & 45.17 & 0.1814 \\
        \hline 
        \cite{Peca2023} & $3 \leq z \leq 3.5$ & 45.79 & 45.78 & 0.1935 \\
        & $3.5 \leq z \leq 4$ & 45.69 & 45.68 & 0.1915 \\
        & $4 \leq z \leq 4.5$ & 45.59 & 45.58 & 0.1894 \\
        & $4.5 \leq z \leq 5$ & 45.48 & 45.48 & 0.1882  \\
        & $5 \leq z \leq 5.5$ & 45.41 & 45.37 & 0.1955 \\
        & $5.5 \leq z \leq 6$ & 45.30 & 45.27 & 0.2096 \\
        \hline
    \end{tabular}
    \caption{The peak X-ray Luminosity, $L_{\rm X}$, and the Gumbel parameters $\alpha$ and $\beta$ for the luminosity functions described in Section \ref{sec:lum_funcs}. }
    \label{tab:full}
\end{table*}


\bsp	
\label{lastpage}
\end{document}